\documentstyle[12pt]{article}
\begin{document}
\renewcommand{\thetable}{\Roman{table}}
\newcommand{\be}{\begin{eqnarray}}
\newcommand{\beq}{\begin{equation}}
\newcommand{\ba}{\begin{array}}
\newcommand{\ee}{\end{eqnarray}}
\newcommand{\eeq}{\end{equation}}
\newcommand{\ea}{\end{array}}
\newcommand{\zt}{\zeta}
\newcommand{\ve}{\varepsilon}
\newcommand{\al}{\alpha}
\newcommand{\gm}{\gamma}
\newcommand{\Gm}{\Gamma}
\newcommand{\om}{\omega}
\newcommand{\et}{\eta}
\newcommand{\bt}{\beta}
\newcommand{\dt}{\delta}
\newcommand{\Dt}{\Delta}
\newcommand{\La}{\Lambda}
\newcommand{\la}{\lambda}
\newcommand{\vp}{\varphi}
\newcommand{\nn}{\nonumber}
\newcommand{\nid}{\noindent}

\begin{titlepage}
\rightline{Preprint of the Institute of Physics of}
\rightline{St. Petersburg State University}
\rightline{SPbU-IP-00-18}
\vspace{0.75cm}
\begin{center}
 {\Large \bf
 Critical thermodynamics of three-dimensional $MN$-component
 field model with cubic anisotropy from higher-loop RG expansions}

\vspace{0.5cm}
 {\Large A. I. Mudrov$^*$,  K. B. Varnashev$^{**}$}

\bigskip
{\it $^{*}$ Institute of Physics, St. Petersburg State University,
Ulyanovskaya 1, Stary Petergof, St. Petersburg, 198904, Russia;
aimudrov@dg2062.spb.edu \\
$^{**}$
Saint Petersburg Electrotechnical University, Professor Popov Street 5, \\
St. Petersburg, 197376, Russia;
kvarnash@kv8100.spb.edu}
\end{center}
\begin{abstract}
\vspace{0.25cm}
The critical behavior of an $MN$-component order parameter Ginzburg-Landau
model with isotropic and cubic interactions describing antiferromagnetic
and structural phase transitions in certain crystals with complicated
ordering is studied in the framework of the four-loop
renormalization group (RG) approach in $(4-2\ve)$-dimensions. Using
dimensional regularization and the minimal subtraction scheme, the
perturbative expansions for RG functions are deduced for generic $M$ and $N$
and resummed by the Borel transformation combined with a conformal mapping.
Investigation of the global structure of RG flows for the physically
significant cases $M=2$ and $N=2, N=3$ shows that the model has a
three-dimensionally stable fixed point different from the
Bose one. The critical dimensionality is proved to be exactly two times
smaller than its counterpart in the real cubic model:
$N_c^C =\frac{1}{2} N_c^R$. The numerical value $N_c^C=1.447 \pm 0.020$ is
obtained from resumming the known five-loop $\ve$-series for $N_c^R$.
Since $N_c^C < 2 $, the critical thermodynamics of the model relevant
to the phase
transitions in real substances should be governed by the complex cubic fixed
point with a new set of critical exponents:
$\gm=1.404(25)$,  $\nu=0.715(10)$, $\et=0.0343(20)$ for $N=2$ and
$\gm=1.390(25)$,  $\nu=0.702(10)$, $\et=0.0345(15)$ for $N=3$.
\end{abstract}

\vspace{0.5cm}
\vspace{1cm}

\nid
To be published in Proceedings of the XXIII-rd Int. Coll.
{\it Group Theoretical Methods in Physics (Group-23)}, JINR RAS, Dubna,
Russia, 2001.

\vspace{1.5cm}

\rightline{{\sl Typeset using} \LaTeX}
\end{titlepage}

\newpage
We study the critical behavior of an $MN$-component field model
with a cubic anisotropy having a number of interesting applications
to phase transitions in three-dimensional simple and complicated
systems. The Landau-Ginzburg-Wilson Hamiltonian of the model reads:
\be
H =
\int d^{~D}x \Bigl[{1 \over 2} \sum_{\al=1}^N (m_0^2~ |\vec \vp^{\al}|^2
 + |\nabla \vec \vp^{\al}|^2)
 + {u_0 \over 4!} \Bigl(\sum_{\al=1}^N |\vec \vp^{\al}|^2 \Bigr)^2
 +  {v_0 \over 4!} \sum_{\al=1}^N |\vec \vp^{\al}|^4 \Bigr],
\label{eq:Ham}
\ee
where $\vp^\al$, for each $\al$, is the $M$-component field
in $D=4 - 2\ve$ dimensions, $m_0$, $u_0$ and $v_0$ are the "bare"
mass and coupling constants, respectively.

For $M=N=2$ Hamiltonian (\ref{eq:Ham}) describes the structural phase
transition in $Nb O_2$ crystal and the antiferromagnetic phase transitions
in $Tb Au_2$ and $Dy C_2$. Another physically important case $M=2$, $N=3$ is
relevant to the antiferromagnetic phase transitions in $K_2 Ir Cl_6$,
 $Tb D_2$ and $Nd$ crystals \cite{Muk}. The magnetic and
structural phase transitions in a cubic crystal are governed by model
(\ref{eq:Ham}) at $M=1$ and $N=3$ \cite{Ah73}.
In the replica limit $N \to 0$ ($M=1$)
Hamiltonian (\ref{eq:Ham}) is known to determine the critical properties of
weakly disordered quenched systems undergoing second-order phase
transitions \cite{GrLutAh} with a specific set of critical exponents
\cite{HLKh}. Finally, the case $M=1$ and $N \to \infty$ corresponds to the
Ising model with equilibrium magnetic impurities \cite{Ah73l}. In this limit
the Ising critical exponents take the Fisher renormalization \cite{F68}.
Since the static critical phenomena in a cubic crystal as well as in randomly
diluted Ising spin systems are well understood 
\cite{MSSh,KSf,KTSf,VKB,CPV,FHY,Vic}, we will focus here on the 
critical behavior 
of the above mentioned multisublattice antiferromagnets. This is the case of 
$M=2$ and $N=2, N=3$ in fluctuation Hamiltonian (\ref{eq:Ham}).

For the first time the magnetic and structural phase transitions in crystals
with complicated ordering described by model (\ref{eq:Ham}) were
studied by Mukamel and Krinsky within the lowest orders in $\ve$ \cite{Muk}.
A new three-dimensionally stable fixed point ("unique" point), different from
the Heisenberg or the Bose one, was predicted. The point was shown to
determine a new universality class with a specific set of critical exponents.
However, for the physically important case $N=2$, the critical exponents of
the unique fixed point turned out to be the same as those of the Heisenberg
one within the two-loop approximation.
For the years an alternative analysis of critical behavior of the model,
RG approach in fixed dimensions, was carried out within the two- and
three-loop approximations \cite{Shp,VS}. Those investigations gave the same
qualitative predictions: the unique stable fixed point does exist on the 3D
RG flow diagram. However, the critical exponents computed at this point with
the use of different resummation procedures were proved to be close to those
of the Bose fixed point rather than the Heisenberg one. It was also shown
that both unique and Bose fixed points
are very close to each other on 3D diagram of RG flows, so that they
may interchange their stability in the next orders of RG
approximations \cite{VS}. Recently critical properties of the model have
been analyzed in third order in $\ve$ \cite{BGMV,MV-1}. Investigation of
the fixed points stability and calculation of the critical dimensionality $N_c$
of the order parameter separating two different regimes of critical behavior
confirmed that the model (\ref{eq:Ham}) possesses the anisotropic stable fixed
point for $N=2$ and $N=3$. By making use of a new approach to summation of
divergent field-theoretical series \cite{MV-2} based on the Borel
transformation in combination with a conformal mapping \cite{LgZ}, the critical
exponents estimates for the unique stable fixed point were obtained
\cite{MV-1}. The values appeared to be close to those of the Heisenberg
point in contradiction to the numerical results given by RG approach directly
in 3D \cite{Shp,VS}. The cause of such a distinction in estimates of the
critical exponents is the quite short, three-loop $\ve$ series used.
So, the aim of this work is to extend
existing three-loop $\ve$ expansions of the model to the four-loop order
and to study critical phenomena in substances of interest more
carefully. Namely, on the basis of the four-loop expansions for RG
functions
obtained in the framework of dimensional regularization and the minimal
subtraction scheme we analyze the stability of the fixed points and
calculate the critical
dimensionality of the order parameter field $N_c^C$. Then we give more
accurate critical exponents estimates applying the summation
approach developed in Ref. \cite{MV-2} to the  four-loop series.

The four-loop $\ve$ expansions for the $\bt$-functions of the model are
as follows
\be
\bt_u &=& 2 \ve u -u^2 - \frac{4}{N + 4} u v
 + \frac{1}{(N+4)^2} \Bigl[3 u^3 (3 N + 7) + 44 u^2 v + 10 u v^2 \Bigr]
\label{eq:Bu} \\
&-& \frac{1}{(N + 4)^3} \Bigl[\frac{u^4}{4}(48 \zt(3) (5 N+11)
 +33 N^2+461 N+740) + u^3 v (384 \zt(3) +79 N
\nn \\
&+& 659) + \frac{ u^2 v^2}{2} (288 \zt(3) +3 N +1078) + 141 u v^3 \Bigr]
 - \frac{1}{(N + 4)^4} \Bigl[\frac{u^5}{12}(- 48 \zt(3) (63 N^2
\nn \\
&+& 382 N + 583) + 144  \zt(4)( 5 N^2+31 N +44) - 480 \zt(5) (4 N^2
 + 55 N + 93)+ 5N^3
\nn \\
&-& 3160 N^2 - 20114 N- 24581) - \frac{2 u^4 v}{3} (12 \zt(3) (3 N^2
 + 276 N + 1214) - 36 \zt(4)
\nn \\
&\times& (19 N + 85) + \zt(5) (2400 N + 23040) - 28 N^2 + 3957 N +15967)
 - \frac{u^3 v^2}{3} (72 \zt(3)
\nn \\
&\times& (19 N + 426) - 4032 \zt(4) + 39840 \zt(5) + 1302 N + 46447)
 + \frac{2 u^2 v^3}{3} (60 \zt(3) (N
\nn \\
&-& 84) - 792 \zt(4) - 4800 \zt(5) - 125 N - 12809)
 - \frac{u v^4}{2} (400 \zt(3) + 768 \zt(4) + 3851) \Bigr]
\nn \\
\bt_v &=& 2 \ve v - \frac{1}{N+4}(6 u v + 5 v^2) + \frac{1}{(N+4)^2}
\Bigl[u^2 v ( 5 N + 41) + 80 u v^2 + 30 v^3 \Bigr]
\label{eq:Bv} \\
&-& \frac{1}{(N + 4)^3} \Bigl[\frac{u^3 v}{2} (96 \zt(3) (N + 7)
 - 13 N^2 + 184 N + 821) + \frac{u^2 v^2}{4} (4032 \zt(3) + 59 N
\nn \\
&+& 5183) + u v^3 (768 \zt(3) + 1093) + \frac{v^4}{2} (384 \zt(3) + 617) \Bigr]
 - \frac{1}{(N + 4)^4} \Bigl[\frac{u^4 v}{4} (48 \zt(3)
\nn \\
&\times& (N^3 -12 N^2 - 140 N - 567) + 144 \zt(4) (2 N^2 + 17 N +45)
-3360 \zt(5) (3 N + 13)
\nn \\
&-& 29 N^3 - 28 N^2 - 6958 N - 19679)
 + \frac{u^3 v^2}{3} (12 \zt(3) (9 N^2 - 591 N -7028) + \zt(4)
\nn \\
&\times& (3528 N + 21240) - 480 \zt(5)(10 N + 287) + 61 N^2 - 5173 N - 66764)
- \frac{u^2 v^3}{3}
\nn \\
&\times& (1800 \zt(3) (N + 62) - 144 \zt(4) (8 N + 203)
 + 172800 \zt(5) + 56 N + 93701)
\nn \\
&-& 4 u v^4 (5090 \zt(3) - 1296 \zt(4) + 7600 \zt(5) + 4503)
 + \frac{v^5}{2} (- 8224 \zt(3) + 1920\zt(4)
\nn \\
&-& 12160 \zt(5)-7975) \Bigr], \nn
\ee
where $\zt(3)$,$\zt(4)$, and $\zt(5)$ are the Riemann $\zt$ functions.

From these equations we find the formal series for the fixed points.
Then we calculate the stability matrix eigenvalues using an approach based
on the Borel transformation
\be
 F(2\ve;a,b) =\sum_{k=0}^\infty A_k(\la) \int_0^\infty
 e^{-\frac{x}{2 a \ve}} \Bigl(\frac{x}{2 a \ve}\Bigr)^b
 d\Bigl(\frac{x}{2 a \ve}\Bigr)
 \frac{\omega^k(x)}{[1-\omega(x)]^{2\la}}.
 \label{eq:Bor}
\ee
modified with a conformal mapping $\omega=\frac{\sqrt{x+1}-1}{\sqrt{x+1}+1}$
\cite{LgZ}, which does not require the knowledge of the exact asymptotic
high-order behavior of the series \cite{MV-2}. If both eigenvalues are
negative, the associated fixed point is infrared stable and the critical
behavior of experimental systems undergoing second-order transitions is
determined only by such a stable point. For the most intriguing the Bose and
the complex cubic fixed points our numerical results are presented in Table I.
It is seen that the complex cubic fixed point is absolutely stable in $D=3$
($2\ve=1$), while the Bose point appears to be of a "saddle-knot" type.
However, $\la_2$'s of either points are very small on the four-loop level,
thus implying that these points may swap their stability in the next order
of RG approximation.

In addition to the eigenvalues, we also calculate the critical dimensionality
$N_c^C$ of the order parameter. The four-loop expansion is
\be
N_c^C = 2 - (2 \ve) + \frac{5}{24} \Bigl[6 \zt(3) -1 \Bigr] (2 \ve)^2
 + \frac{1}{144} \Bigl[45 \zt(3) + 135 \zt(4) - 600 \zt(5) -1 \Bigr] (2 \ve)^3.
\label{eq:Nc}
\ee
Instead of processing this expression numerically,
we establish the exact relation $N_c^C=\frac{1}{2} N_c^R$, which is
independent on the order of approximation used. Here $N_c^C$ and $N_c^R$ are
the critical dimensionalities in the complex and in the real cubic model,
respectively.
The five-loop $\ve$-expansion for $N_c^R$ was obtained in Ref. \cite{KSf}.
Resummation of that series gave the estimate $N_c^R=2.894(40)$ \cite{VKB}.
Therefore we conclude that $N_c^C=1.447(20)$ from the five-loops.
Practically the same estimate $N_c^C=1.435(25)$ follows from a 
constrained analysis of $N_c^R$ that takes into account that in two 
dimensions $N_c^R=2$ \cite{CPV}. So, the phase transitions in the $Nb O_2$ 
crystal and in the antiferromagnets $Tb Au_2$,
$Dy C_2$, $K_2 Ir Cl_6$, $Tb D_2$, and $Nd$ are of second order and their
critical thermodynamics should be controlled by the complex cubic fixed point
with a specific set of critical exponents, in the frame of given approximation.
Corresponding four-loop critical exponents estimates are displayed in Table II.

Although our results seem to be self-consistent, still there is definite
contradiction with the nonperturbative theoretical predictions \cite{CB}.
On can hope, however, that the five-loop contributions being taken into
account will eliminate this inconsistency.

The authors are grateful to Dr. E. Blagoeva for sending a copy of her
important work.

\newpage
\begin{table}
\caption{Eigenvalue exponents estimates obtained for the Bose (BFP) and the
complex cubic (CCFP) fixed points at $N=2$ and $N=3$ within the
four-loop approximation in $\ve$ ($2\ve=1$) using Borel transformation with
a conformal mapping.}
\label{TabI}
\vspace{0.5cm}
\hspace{1.5cm}
\begin{tabular}{|c|l|l|l|l|}\hline
Type of             &\multicolumn{2}{|c|}{$N=2$}&
                     \multicolumn{2}{|c|}{$N=3$}
                   \\[0pt]    \cline{2-5}
fixed point         &\multicolumn{1}{|c|}{$\la_1$}
                    &\multicolumn{1}{|c|}{$\la_2$}
                    &\multicolumn{1}{|c|}{$\la_1$}
                    &\multicolumn{1}{|c|}{$\la_2$}
                   \\[0pt] \hline
BFP                 &$-0.395(25)$ & $0.004(5)$
                    &$-0.395(25)$ & $0.004(5)$
                   \\[0pt] \hline
CCFP                &$-0.392(30)$ & $-0.030(10)$
                    &$-0.400(30)$ & $-0.015(6)$
                   \\[0pt]\hline
\end{tabular}
\end{table}

\begin{table}
\caption{Critical exponents calculated for the Heisenberg (HFP), the Bose (BFP)
and the complex cubic (CCFP) fixed points at $N=2$ and $N=3$ within the
four-loop approximation in $\ve$ ($2\ve=1$) using Borel transformation with a conformal
mapping.}
\label{TabII}
\vspace{0.5cm}
\begin{tabular}{|c|l|l|l|l|l|l|}\hline
Type of             &\multicolumn{3}{|c|}{$N=2$}&
                     \multicolumn{3}{|c|}{$N=3$}
                   \\[0pt]    \cline{2-7}
fixed point         &\multicolumn{1}{|c|}{$\eta$}
                    &\multicolumn{1}{|c|}{$\nu$}
                    &\multicolumn{1}{|c|}{$\gm$}
                    &\multicolumn{1}{|c|}{$\eta$}
                    &\multicolumn{1}{|c|}{$\nu$}
                    &\multicolumn{1}{|c|}{$\gm$}
                   \\[0pt]  \hline
HFP                 &$0.0343(15)$ & $0.725(15)$ & $1.429(20)$
                    &$0.0317(10)$ & $0.775(15)$ & $1.524(25)$
                   \\[0pt] \hline
BFP                 &$0.0348(10)$ & $0.664(7)$ & $1.309(10)$
                    &$0.0348(10)$ & $0.664(7)$ & $1.309(10)$
                   \\[0pt] \hline
CCFP                &$0.0343(20)$ & $0.715(10)$ & $1.404(25)$
                    &$0.0345(15)$ & $0.702(10)$ & $1.390(25)$
                   \\[0pt]\hline
\end{tabular}
\end{table}

\nid

\end{document}